\documentclass[pra,twocolumn,showpacs,letterpaper,showpacs,superscriptaddress, nofootinbib]{revtex4-1}
\usepackage{graphicx,amsmath,amssymb,amsfonts,latexsym,color,dcolumn,bm}
\usepackage{braket}
\usepackage{makeidx}
\usepackage{multirow}
\usepackage[dvipsnames,svgnames,table]{xcolor}
\usepackage{graphicx}
\usepackage{epstopdf}
\usepackage{ulem}
\usepackage{here}

\begin{document}
\newcommand{\be}{\begin{equation}}
\newcommand{\ee}{\end{equation}}
\newcommand{\bea}{\begin{eqnarray}}
\newcommand{\eea}{\end{eqnarray}}
\newcommand{\ad}{a^{\dag}}
\newcommand{\la}{\langle}
\newcommand{\ra}{\rangle}
\newcommand{\om}{\omega}
\newcommand{\Ep}{E^{(+)}}
\newcommand{\Em}{E^{(-)}}
\newcommand{\pa}{\partial}

\title{The photon, its mode function and complementarity}
\author{R. Menzel}
\affiliation{Institut f\"ur Physik und Astronomie, Universit\"at Potsdam, D14476 Potsdam, Germany}
\author{R. Marx}
\affiliation{Institut f\"ur Physik und Astronomie, Universit\"at Potsdam, D14476 Potsdam, Germany}
\author{D. Puhlmann}
\affiliation{Institut f\"ur Physik und Astronomie, Universit\"at Potsdam, D14476 Potsdam, Germany}
\author{A. Heuer}
\affiliation{Institut f\"ur Physik und Astronomie, Universit\"at Potsdam, D14476 Potsdam, Germany}
\author{W. P. Schleich}
\affiliation{Institut f\"ur Quantenphysik and Center for Integrated Quantum Science and Technology (IQ$^{ST}$), Universit\"at Ulm, Albert-Einstein-Allee 11, D-89081 Ulm, Germany} 
\affiliation{Hagler Institute of Advanced Study at Texas A\&M University, Texas A\&M AgriLife Research, Institute of Quantum Science and Engineering (IQSE), and Department of Physics and Astronomy, Texas A\&M University, College Station TX 77843-3572,USA} 
\date{}

\begin{abstract}

We probe the principle of complementarity by performing a double-slit experiment based on entangled photons created by spontaneous parametric down-conversion from a pump mode in a TEM$_{01}$-mode. Our setup brings out the need for a careful selection of the signal-idler photon pairs for our study of visibility and distinguishability. Indeed, when the signal photons interfering at the double-slit belong to this double-hump mode we obtain almost perfect visibility of the interference fringes and no ``which-slit'' information is available. However, when we break the symmetry between the two maxima of the mode by detecting the entangled idler photon, the paths through the slits become distinguishable and the visibility vanishes. It is the mode function of the photons selected by the detection system which decides if interference, or ``which-slit'' information is accessible in the experiment.
\end{abstract}



\maketitle

\section{Introduction}

Photons \cite{Fermi32} are discrete excitations of mode functions of the classical electromagnetic field \cite{Lamb95} and reveal themselves in excitations of a detector \cite{Glauber07} such as an atom. The principle of complementarity \cite{Bohr28} of quantum theory \cite{Bohm89} can give rise to mind-boggling phenomena \cite{Schleich16} in single-photon interference experiments ranging from the quantum eraser \cite{Scully82, Scully91, Kim00} via induced coherence \cite{Zou91,  Heuer15} to delayed-choice experiments \cite{Ma16}. In the present article we report on a double-slit experiment with entangled photons designed to bring out most clearly the intimate connection between the mode function of a single photon and complementarity. 

\subsection{Complementary variables}

The year 2018 marks the 90th anniversary of the publication of Niels Bohr`s article \cite{Bohr28} ushering in the principle of complementarity. His insight into the inner workings of quantum theory was guided by the observation that:

``...the measurement of the positional coordinates of a particle is accompanied not only by a finite change in the dynamical variables, but also the fixation of its position means a complete rupture in the causal description of its dynamical behavior, while the determination of its momentum always implies a gap in the knowledge of its spatial propagation. Just this situation brings out most strikingly the complementary character of the description of atomic phenomena which appears as an inevitable consequence of the contrast between the quantum postulate and the distinction between object and agency of measurement, inherent in our very idea of observation.''

Notwithstanding the fact that Bohr associates with the act of a measurement a back action \cite{Storey94} on the system that is measured, he clearly identifies complementary variables such as position and momentum of a particle.

\begin{figure*} [t]
\includegraphics[width=1\linewidth]{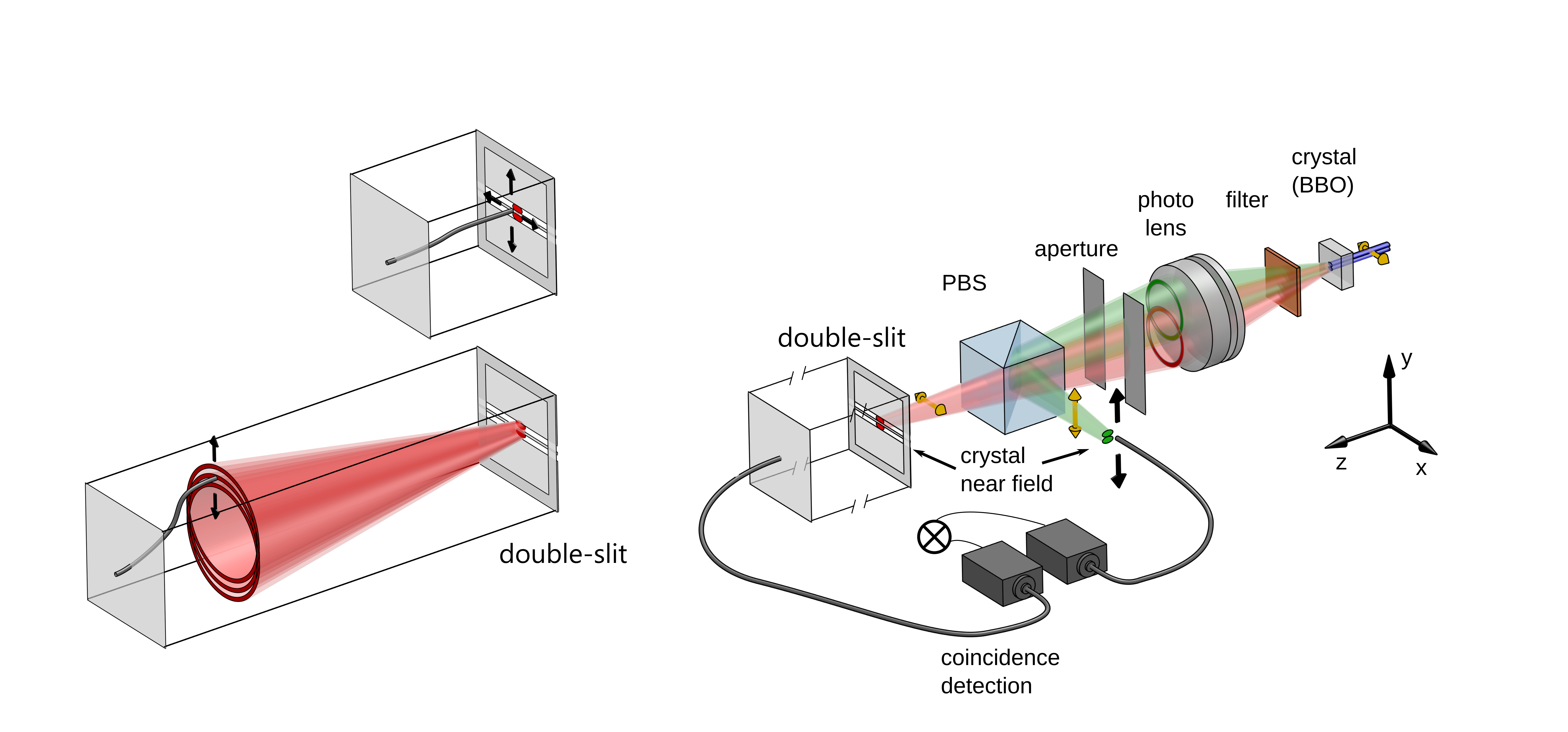}
\caption{``Which-slit'' information and interference in a double-slit experiment based on coincidence measurements of entangled photons (right). The photons are created by spontaneous parametric down-conversion (SPDC) due to a light field in a coherent double-hump mode pumping a BBO-crystal. A polarizing beam splitter (PBS) separates the idler and signal photons. The signal photons pass through a double-slit, and are detected either in its near-field (top left), or far-field (bottom left). The idler photons outcoupled at the PBS do not traverse the double-slit and are measured in coincidence with the signal photons at a position orthogonal to the optical axis which has the same distance from the PBS as the double-slit. In contrast to our earlier setup \cite{Menzel12, Menzel13} we now have apertures in the far-field of the light cones between the photo lens and the PBS. They select only portions of the emitted bi-photons before obtaining ``which-slit'' information (top left) or visibility (bottom left) of the signal photon in the near or far field of the double-slit, respectively. Although we here only depict the example of a slit-aperture we also employ circular apertures.}\label{fig1}
\end{figure*}
His subsequent dialogue \cite{Wheeler84} with Albert Einstein centered around the double-slit experiment where ``which-path'' information and interference play the role of complementary variables \cite{Hellmuth87}. In particular, Bohr argued \cite{Wheeler84} that quantum mechanics does not allow us to determine through  which slit the particle traversed, and at the same time observe the resulting interference pattern in the far field. His conviction is rooted in the necessity to have two mutually exclusive experimental setups to obtain ``which-path'' information and interference. This requirement also reflects the complementary views of particle and wave nature of matter.

We emphasize that considerations of this kind focus exclusively on the use of a single photon. Indeed, a recent version \cite{Menzel12, Bolduc14} of a double-slit experiment employed a pair of entangled photons created by spontaneous parametric down-conversion (SPDC) \cite{Reintjes84}. Hence, one of the two photons, such as the idler photon, can provide us with the path information, while the other, that is the signal photon, yields interference as shown in Fig.1. In this arrangement the idler photon, in contrast to the signal photon, does not even pass the double-slit \cite{footnote01}. 

Despite the apparent ``which-path'' information encoded in the idler photon the interference pattern of the signal photons recorded in coincidence with the idler photon displays a remarkably large contrast. Nevertheless, these on first sight startling observations are not in conflict with the principle of complementarity as emphasized in Ref. \cite{Schleich16, Menzel12, Menzel13}.

A crucial ingredient of the setup of Fig.1 is the shape of the pump mode with two maxima in the transversal electric field distribution and a node between them. This unusual pump mode is transferred onto the signal and idler modes. Only when the two maxima match the slits can we get ``which-path'' information. Indeed, here we take advantage of the fact that the signal and idler photons are always in the same maximum as demonstrated in Ref. \cite{Menzel12, Menzel13}.

So far we have focused exclusively on the near field properties of the modes but it is the far-field that determines the interference properties. In the present article we show that in the presence of such a pump mode the far-fields of the signal and idler photons display a remarkable asymmetry in their familiar ring structure. The top and the bottom of the ring belong to two different modes with either one or two maxima. This property has immediate consequences on our double-slit experiment with entangled photons \cite{Menzel12, Menzel13, Elsner15} and, in particular, on the principle of complementarity.

\subsection{Outline}

Our article is organized as follows. In Sec. II we formulate the problem of obtaining ``which-path'' information \textit{and} observing interference in our double-slit experiment based on entangled photons. Here we emphasize the crucial role of pumping the crystal in TEM$_{01}$ mode. We then in Sec. III briefly summarize the experimental setup and emphasize crucial differences to earlier versions \cite{Menzel12, Menzel13} resulting from additional apertures. In particular, we include an experimental and a numerical study on the influence of a slit-aperture on the visibility and distinguishability. We then in Sec. IV point out a surprising asymmetry between the top and the bottom of the light cones formed by the photons. This effect originates from the phase matching condition and the unusual pump mode. Since our experimental results are in complete agreement with the corresponding numerical simulations \cite{Elsner15} we can identify the signal-idler pairs belonging to the same mode. Only these photons have to obey the principle of complementarity as discussed in Sec. V. Finally we conclude in Sec. VI by summarizing our results.

\section{Formulation of Problem}

In our previous experiments \cite{Menzel12, Menzel13} we have used SPDC to obtain ``which-path''  information and interference in a double-slit experiment. For this purpose we have pumped a crystal by a TEM$_{01}$-mode and have imaged the two intensity humps of the down-converted light onto a double-slit in the near field of the crystal as shown in Fig.1. In this way we not only observe the interference pattern in the far field but also obtain ``which-slit'' information from the coincidence measurement of the two entangled photons in the near field. However, in sharp contrast to a naive application of the principle of complementarity we have found \cite{Menzel12, Menzel13} a high visibility in the far-field interference pattern of the signal photon even when observed in coincidence with the near-field idler photon.

A valuable hint towards an explanation of this surprising phenomenon was given in Ref. \cite{Bolduc14} arguing that the photons providing us with ``which-slit'' information, and those which contribute to the interference pattern \cite{Leach16} belong partially to different photon pairs. This picture is consistent with our earlier observation \cite{Schleich16, Menzel12, Menzel13} based on a rather elementary model using a frozen gas of atoms instead of the crystal with its phase matching conditions. Indeed, the atoms creating the entangled photon pair, and providing us with ``which-slit'' information in the near field are different \cite{Schleich16} from those creating the interference in the far field. The principle of complementarity ensures that a single atom cannot yield ``which-path'' information \textit{and} interference. More detailed calculations \cite{Happ15, Happ16} have reconfirmed this separation into distinct ensembles of atoms.

Although these arguments demonstrate already that the observations reported in Ref. \cite{Menzel12, Menzel13} need not to be in contradiction with the principle of complementarity, they do not bring out the underlying workings of the experiment. In the present article we fill this gap and show that the phase matching condition \cite{Reintjes84} for the nonlinear crystal pumped by a TEM$_{01}$-mode creates different interference patterns in the upper and lower portions of the light cones.

Indeed, the upper part reflects the TEM$_{01}$-mode structure of the single photons at the double-slit, while the lower one is of the TEM$_{00}$-type. This distinction can even be made by the bear eye since the fringes on the top of the ring show a minimum in the middle reflecting the phase shift of $\pi$ between the two humps of the field at the double-slit. 

Due to these subtleties in the modes our earlier experiments \cite{Menzel12, Menzel13} selected only a small share of the measured photons in agreement with the fair-sampling criticism of Ref. \cite{Bolduc14, Leach16}. However, the detailed investigations reported in the present article not only put to rest the remaining questions associated with this discussion but also demonstrate the feasibility of the main idea of \cite{Menzel12, Menzel13} to employ the TEM$_{01}$-pump mode in a test of complementarity.

In particular, our analysis confirms that a single photon in a TEM$_{01}$-like mode structure generated via a thermal SPDC process can produce interference in a double-slit experiment. Photons belonging to the same mode interfere with a high visibility V.

However, if the detection system selects photons with increasing ``which-slit'' information, that is distinguishability D, they do not belong to the same mode anymore and V decreases. This behavior is the manifestation of complementarity in the spatial domain, demonstrated here for a higher-order Gauss-Laguerre mode.

We emphasize that in our experimental arrangement we always satisfy the familiar inequality \cite{Greenberger88, Englert96, Quian}
\begin{equation}
V^2 + D^2 \leq 1 \label{eq1}
\end{equation}
and thereby confirm the principle of complementarity.

\begin{figure*}[t]
\includegraphics[width=1\linewidth]{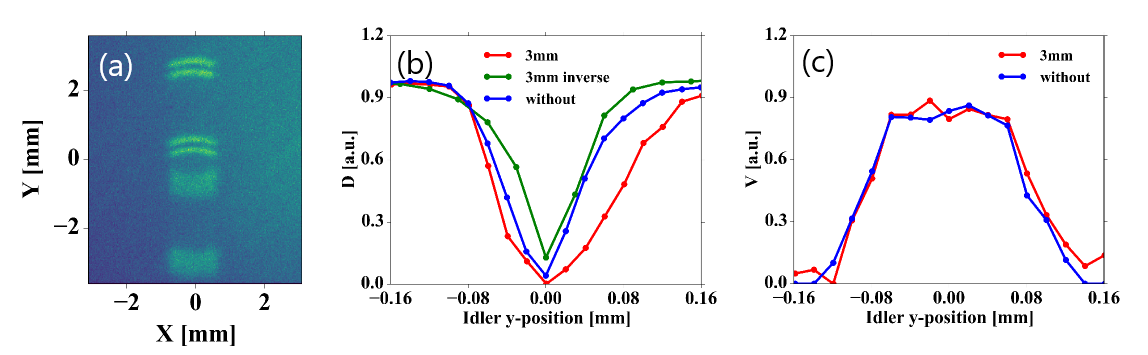}
\caption{Influence of a vertical slit-aperture on the visibility V and distinguishability D defined by Eqs.(2) and (3) in the double-slit experiment with entangled photons. Single-photon interference in the far field (a) without the polarizing beam splitter (PBS) but with a vertical slit-aperture between the crystal and the double-slit as depicted in Fig.1. The residual upper and lower rings (a) with only 30\% of all photons belong to the idler and signal photons. The distinguishability D corresponding to ``which-slit'' knowledge (b) measured in the presence of the PBS for the signal photons behind the double-slit as a function of the position of the idler detector depends sensitively on the width of the slit-aperture as illustrated for the three cases of no aperture, one of width 3mm and an inverse slit-aperture (blocking the middle part) of 3mm width. In contrast, the visibility (c) of the signal photons in the far field is almost independent of it.}\label{fig2}
\end{figure*}

\section{A test of complementarity}

In this section we describe our double-slit experiment with entangled photons \cite{Menzel12} and outline our measurement strategy to deduce visibility and distinguishability. Moreover, we present experimental and numerical results on the influence of a slit-aperture on these two quantities central to a test of the principle of complementarity based on the inequality, Eq.(1).

\subsection{Experimental setup}

Our experimental setup shown in Fig.1 is discussed in more detail in Ref. \cite{Menzel12}. However, in order to keep our article self-contained, we now briefly summarize the essential ingredients.

The initial TEM$_{00}$-mode of the pump laser (Toptica, Blue Mode) with a cw-power of 30 mW and wavelength of 405 nm propagating along the z-axis was converted to a TEM$_{01}$-mode using a phase plate and the spot size at the crystal was 150 $\mu$m. The BBO-crystal cut for type-II phase matching had  a length of 2 mm \cite{footnote02}, and the slits of width 65 $\mu$m were separated by 235 $\mu$m. The detectors were Perkin Elmer (AQR14) fiber-coupled single-photon avalanche photodiodes.

\begin{figure*}
\includegraphics[width=0.8\linewidth]{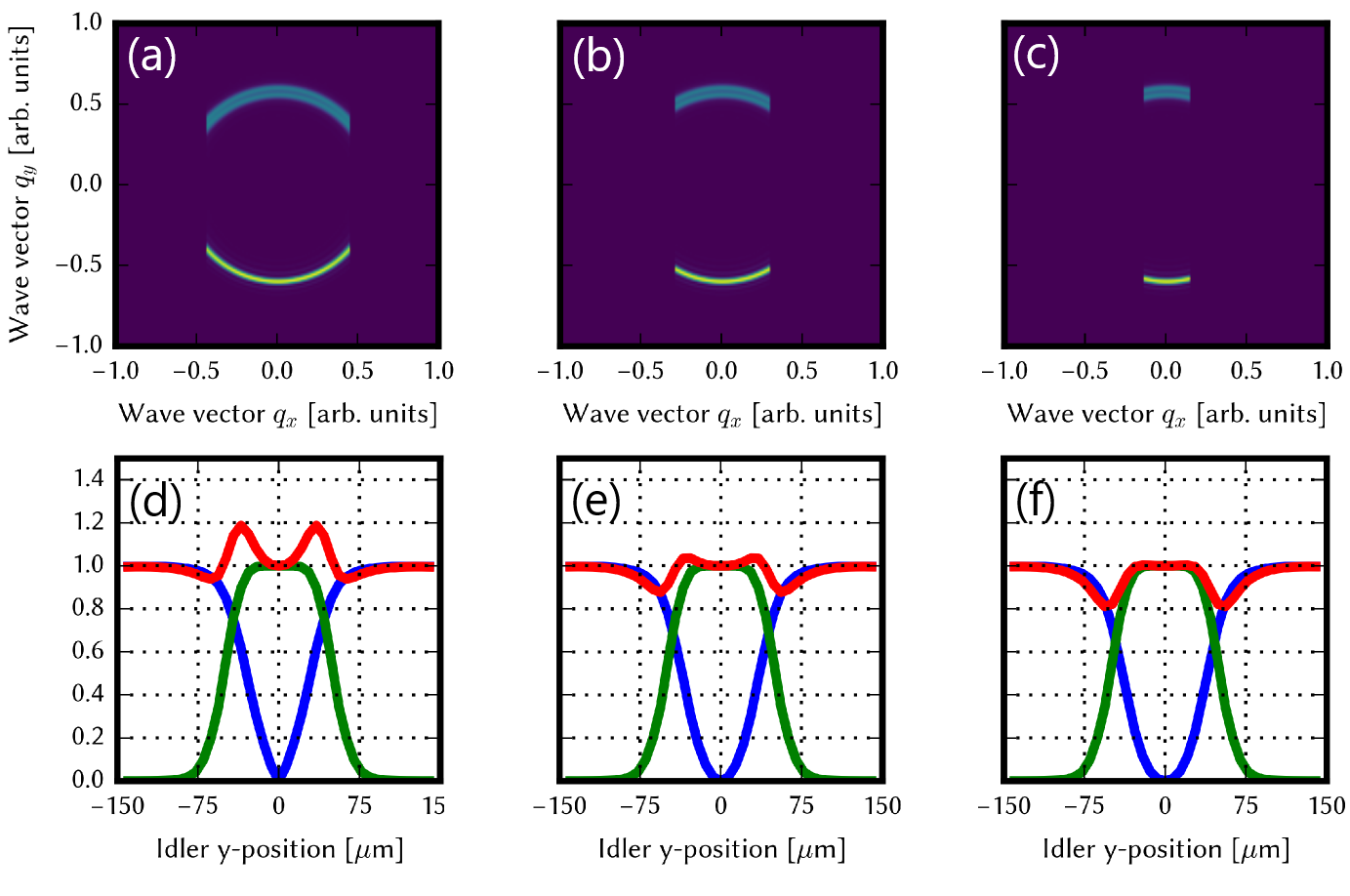}
\caption{Influence of the slit width of a vertical slit-aperture on the visibility V (green) and distinguishability D (blue), as well as the sum (red) $V^2+D^2$  analyzed by a numerical simulation of our experiment. In the top row we depict the signal photons corresponding to 80\% (a), 50\% (b) and 23\% (c) of the original cone diameter observed in the far-field, that is as a function of the transverse wave vector components q$_x$ and q$_y$. The remaining signal photons give rise to values of V and D that lead to a violation of the inequality Eq.(1) at \textit{some} positions of the idler detector (d) and (e), but also to a situation where this condition is satisfied for \textit{all} locations (f). The scales on the corresponding axes are identical in each row.}\label{fig3}
\end{figure*}

Since the two light cones formed by the emitted signal and idler photons have different polarizations we can separate them by a polarizing beam splitter (PBS). We measure the signal photons after a double-slit located in the near field of the crystal which is the imaging plane of its end surface, and aligned parallel to the x-y plane with slits along the x-axis. This arrangement is indicated on the top left of Fig.1.

In another one we use an imaging system to place the signal detector in the far field of the crystal and observe in this way the interference pattern of the signal photons due to the double-slit. Indeed, by moving this detector along the vertical direction as shown in Fig.1 on the bottom left we deduce the visibility
\begin{equation}
V \equiv \frac{R_{max} - R_{min}}{R_{max} + R_{min}} \label{eq2}
\end{equation}
of the interference fringes from the maximal and minimal single photon count rates R$_{max}$ and R$_{min}$.

The idler detector positioned perpendicular to the signal path behind the beam splitter in the near field of the crystal can also be moved in the vertical direction which corresponds to the detection of the idler photons at the position of the double-slit. In this way we obtain ``which-slit'' information for the signal photons by taking advantage of their entanglement with the idler photons.

The distinguishability
\begin{equation}
D \equiv \frac{C_{S1} - C_{S2}}{C_{S1} + C_{S2}} \label{eq3}
\end{equation}
between the two paths 1 and 2 of the signal photons follows from the coincidence count rates C$_{S1}$ and C$_{S2}$ of the signal photons in path 1 or 2 with respect to the idler photons in path 1.

\subsection{Influence of slit-aperture}

In contrast to our earlier experiments \cite{Menzel12, Menzel13} we now insert between the photo lens and the beam splitter a slit-aperture as shown in Fig.1. This tool allows us to select in the far field of the emitted light cones specific areas of the rings appearing on the detection screen cutting the cones along the x-y plane. In this way we use only a small portion of the ring in the determination of the visibility of the signal photons and restrict ourselves in the ``which-slit'' information to those idler photons that belong to the measured signal photons. 

We start by analyzing V and D as a function of the vertical position of the idler detector in the presence of the slit-aperture indicated in Fig.1. In this arrangement mainly the middle portions of the rings corresponding to signal and idler photons survive as exemplified by Fig.2a.

In contrast to the ``which-slit'' knowledge D shown in Fig.2b which sensitively depends on the slit width of the slit-aperture, the visibility V displayed in Fig.2c is almost independent of it. The latter result is not surprising because this part of the far field is not influenced by the slit-aperture. The corresponding decrease in D is in agreement with the fair sampling argument of Ref. \cite{Bolduc14}.

Without the slit-aperture a maximum value of $V^2 + D^2 \approx 1.4$ was determined experimentally \cite{Bolduc14} and theoretically \cite{Elsner15} as a consequence of the unfair sampling. However, even in the presence of the vertical slit-aperture violations of the inequality, Eq.(1), occur in specific domains of the position of the idler detector as confirmed by our simulations \cite{Elsner15}.

Indeed, in the top row of Fig.3 we depict for three different widths of the slit (a-c) the remaining signal photons. In the bottom row (d-f) we study the corresponding dependence of D, V and the sum $V^2+D^2$ on the idler position. For a decreasing width the curve (blue) associated with D is broadened, whereas the corresponding one (green) for V is almost unchanged. Only for the narrowest slit-aperture (c) and (f) the value (red) of the sum $V^2+D^2$ is for all positions of the idler detector smaller or equal to unity. Therefore, our simulations confirm the idea \cite{Bolduc14} of a biased sampling.

\begin{figure} [h]
\includegraphics[width=0.7\linewidth]{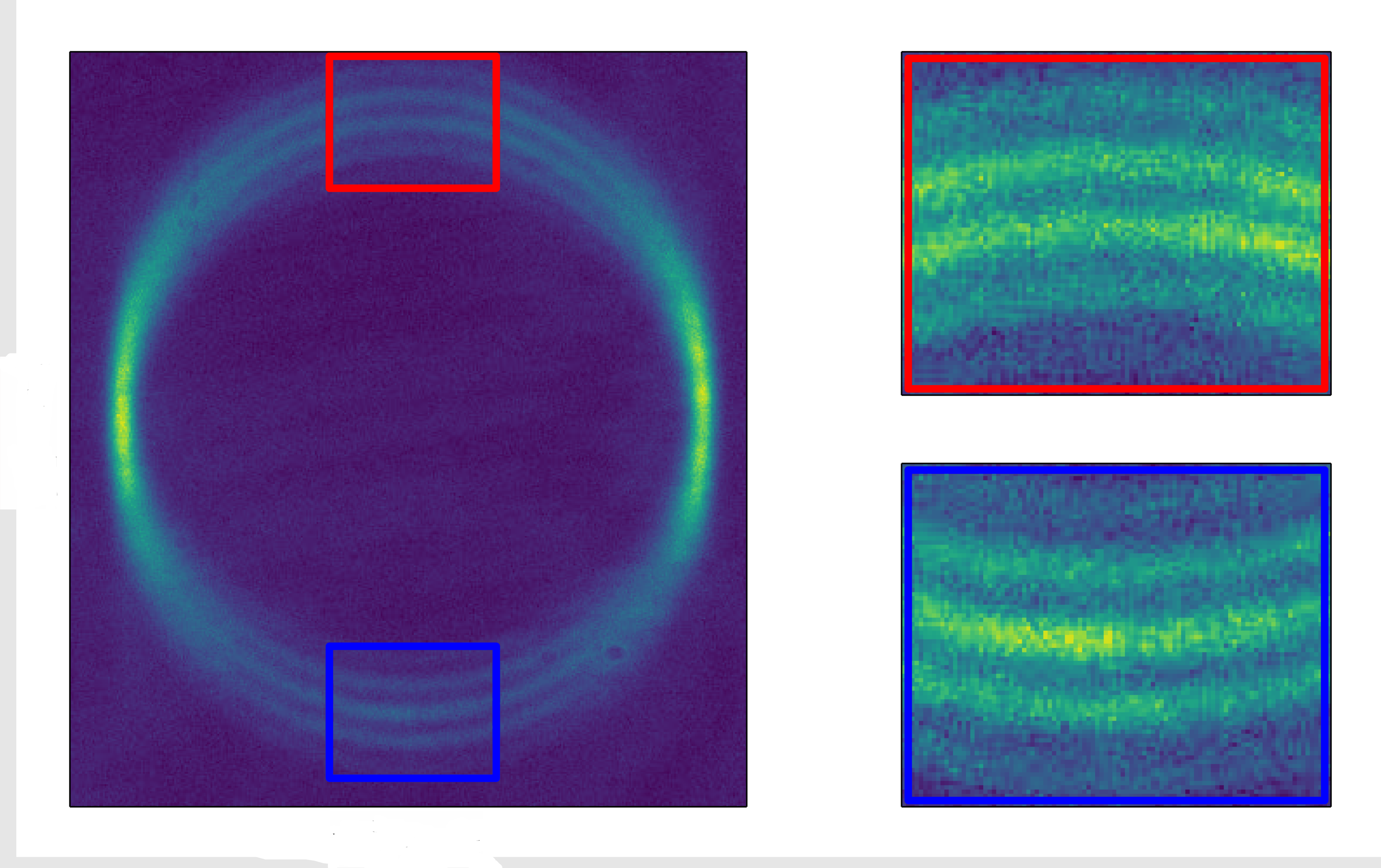}
\caption{Asymmetry in the far-field ring pattern (left) of the measured signal photons generated by SPDC using a type-II crystal pumped by a TEM$_{01}$-mode and having propagated through a double-slit. The upper and lower parts show an even (top right) and an odd (bottom right) number of interference fringes, respectively. Here only signal photons were recorded.}\label{fig4}
\end{figure}

\begin{figure*} [t]
\includegraphics[width=0.9\linewidth]{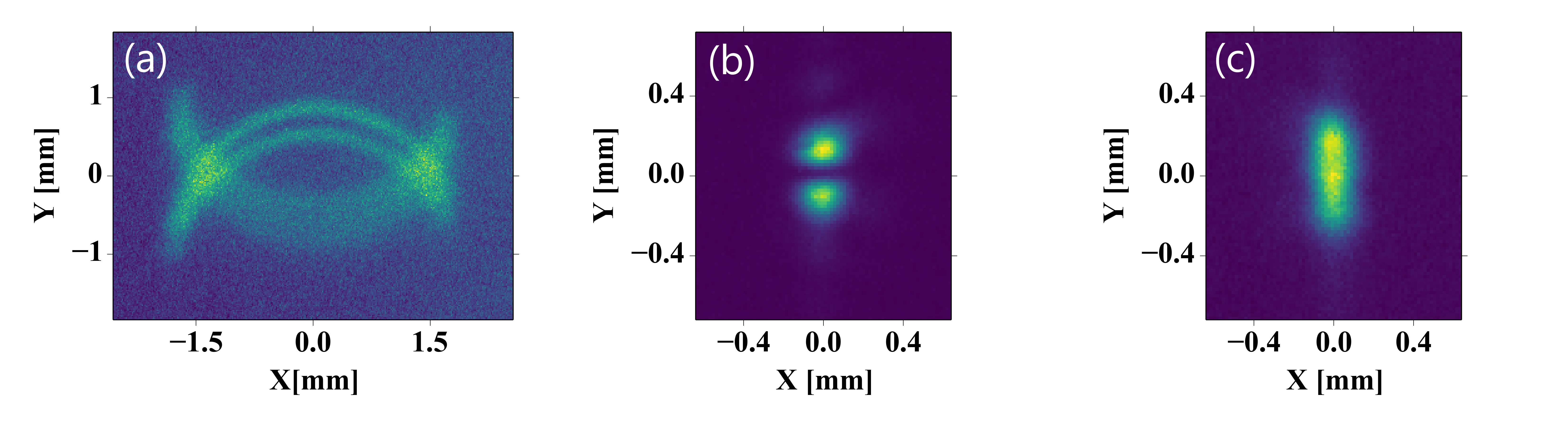}
\caption{Asymmetry in the upper and lower parts of the ring identified as different modes. In contrast to Fig.4 we have inserted now a circular aperture located between the crystal and the polarizing beam splitter (PBS). Applying the circular aperture in the upper part of the ring with an {\it even} number of substructures appearing in the far field (a) results in a {\it double-hump} intensity distribution at the slit (b), and applying the aperture in the {\it lower part} of the ring with an odd number of substructures (a) results in a {\it non-modulated} distribution at the slit (c). For simplicity both signal and idler photons are measured in the left picture (a) showing identical but displaced far-field ring structures.}\label{fig5}
\end{figure*}

\section{Asymmetry of Rings}

In the preceding section we have shown that a slit-aperture reduces the violation of the inequality, Eq.(1), but does not eliminate it altogether. We now identify the deeper origin for this phenomenon by demonstrating experimentally and theoretically that pumping the crystal in the TEM$_{01}$-mode produces a more sophisticated geometrical structure in type-II down-conversion than in type-I which was central to the investigation of Ref. \cite{Bolduc14}. In particular, it leads to a superposition of modes at the location of the double-slit which is at the very heart of our discussion of the principle of complementarity.

\subsection{Different modes on top and bottom}

With our crystal of length 2 mm, and its optical axis at an angle of 41.9 degree with respect to the y-direction, two overlapping rings corresponding to the signal and the idler photons form in the far field. We emphasize that these rings are not rotationally symmetric.

Indeed, Fig.4 shows that the light cone of the signal photons in the far-field displays an even and an odd interference pattern on the top and the bottom of the ring, respectively. In order to bring this difference out most clearly we have enlarged these structures in the right column of Fig.4 and emphasize that they are perfectly reproduced in the simulations of Ref. \cite{Elsner15}.

We study this asymmetry by positioning small apertures in the far field between the crystal and the double- slit for the lower and upper parts of the rings. Figure 5 shows that the even substructure in the far field (top structure  in Fig.5a) belongs to the double-hump intensity distribution (b) at the double-slit position in the near field, and the odd substructure (bottom structure Fig.5a) is produced by a non-modulated mode at the double-slit (c).

In the latter case we have almost a TEM$_{00}$-like near-field distribution covering both slits instead of the expected TEM$_{01}$-mode. Only the upper part with the even interference pattern gives rise to nicely-separated spots reminiscent of the TEM$_{01}$-mode designed to fit the slit. Indeed, the dip in the middle of the far-field interference pattern behind the slit depicted on the top of Fig.4 is a proof of the phase delay of $\pi$ between the two humps. As shown in Ref. \cite{Menzel12} the photons in this type of mode are in a superposition of two wave vectors giving rise to the familiar far-field interference pattern consisting of two maxima and a zero between them.

\subsection{Mode matching}

The upper and the lower parts of the rings correspond to different modes. We now show that they originate from the fact that the phase matching condition in the crystal for a TEM$_{01}$-pump mode results in more complex structures of the down-converted light compared to the commonly used TEM$_{00}$-mode pumping.

To bring this fact out most clearly we first study the signal photons in the \textit{absence} of the double-slit based on a numerical simulation \cite{Elsner15}. Figure 6 shows that even in this situation the asymmetry in the ring prevails.

\begin{figure*}
\includegraphics[width=0.7\linewidth]{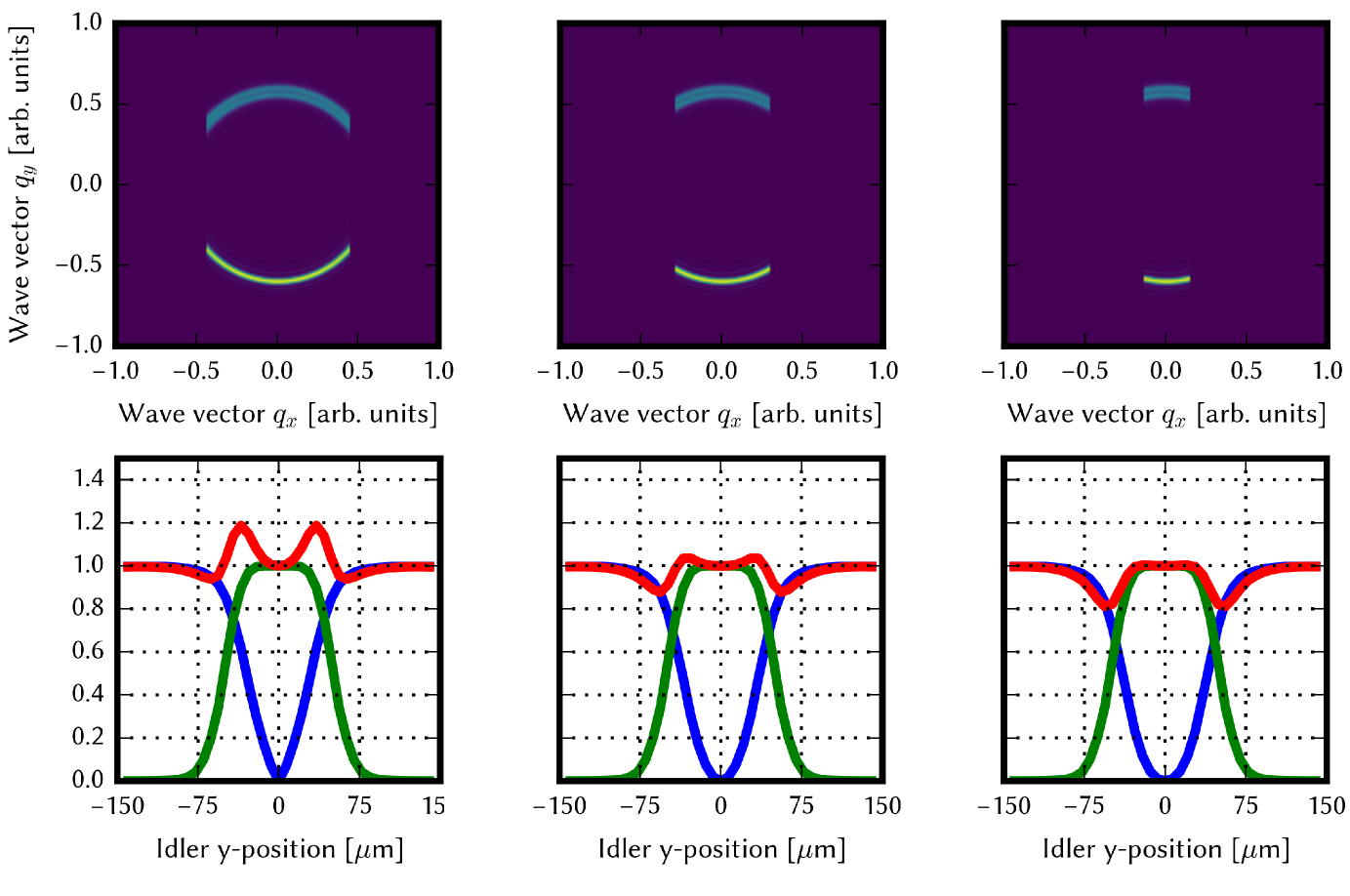}
\caption{Asymmetry in the ring obtained by a numerical simulation of the signal photons generated by SPDC using a type-II crystal pumped by a TEM$_{01}$-mode, and observed in the far field without the double-slit. Whereas the lower part displays a single intensity maximum, the upper one clearly enjoys a double-ring structure. }\label{fig6}
\end{figure*}

In order to understand the influence of this behavior on the coincidence measurements we decompose the bi-photon amplitude distribution
\begin{equation}
\Phi = u({\bf q}_s +{\bf q}_i) sinc(\Delta k_z({\bf q}_s,{\bf q}_i)L/2) \label{eq4}
\end{equation}
into the pump spectral condition and the requirement of phase matching. Here $u$ denotes the transverse mode function of the pump with the wave vectors  ${\bf q}_s$ and ${\bf q}_i$ of the signal and idler photon. Moreover, the type-II crystal of length L leads to a phase miss-match $\Delta k_z({\bf q}_s,{\bf q}_i)L/2$.

In the far field the components q$_{sy}$ and q$_{iy}$ are linearly related to the spatial coordinates as shown in Fig.7 by the double-hump structured straight line determined by $u$. However, the pump condition results in the hyperbolic curve. Only at the intersections of these two qualitatively different curves do we obtain correlated photons.

\begin{figure}
\includegraphics[width=0.7\linewidth]{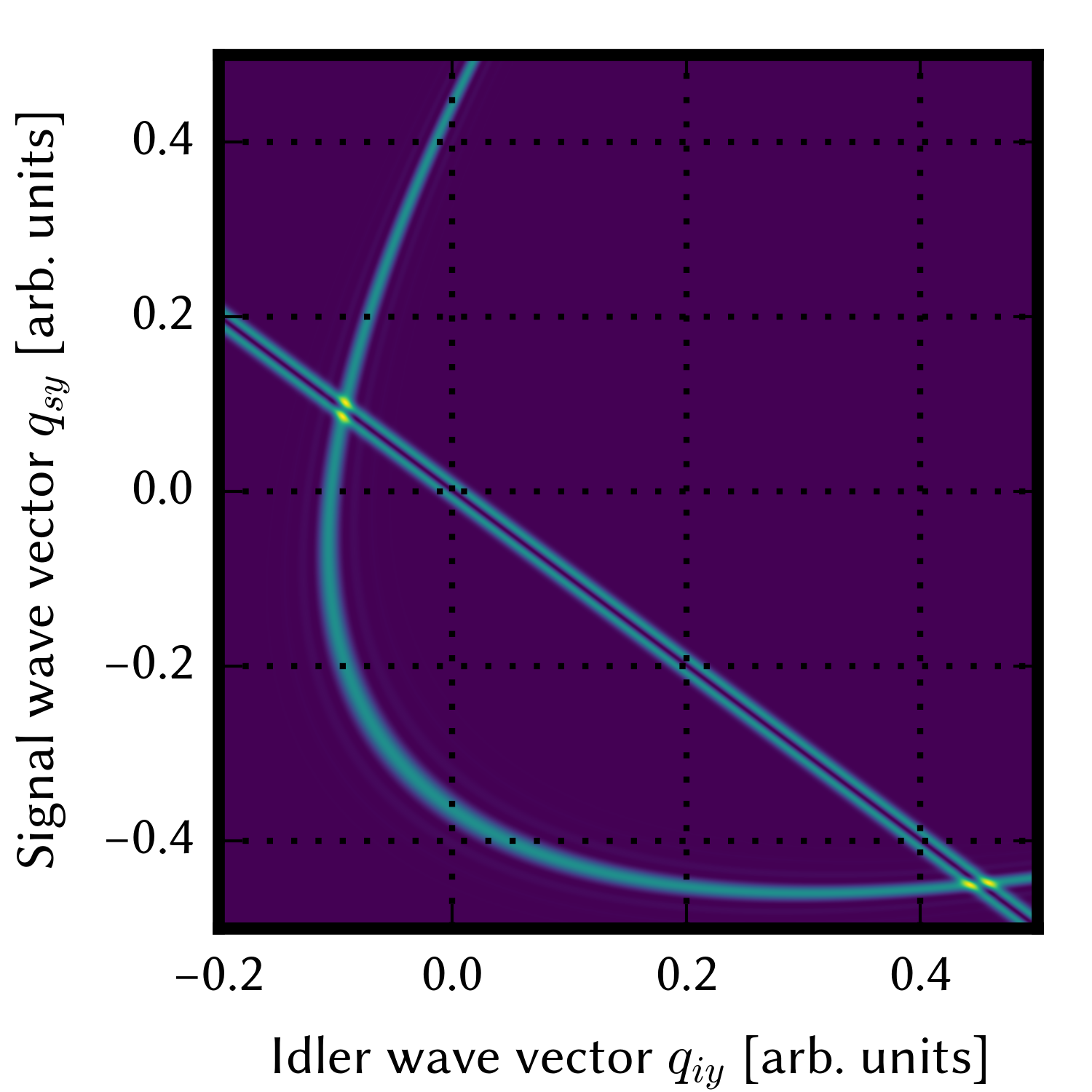}
\caption{Origin of the different interference patterns appearing in the top and bottom parts of the ring shown in Fig.6 identified as energy-momentum conservation required for phase matching. This condition results in transverse wave vector components q$_{sy}$ and q$_{iy}$ which are related to each other in a linear way and depicted here for the two pump humps by straight lines. Energy conservation at the pump and phase matching give rise to a spectral restriction on the two photons which creates the two hyperbolic curves. The two resulting intersections which are qualitatively different -- one rather steep and one almost flat -- provide us with the emission directions of the two photons.}\label{fig7}
\end{figure}

The count rate of the signal photons follow from a projection of the coincidence count rate of Fig.7 onto the corresponding axis. As a consequence, we find in the upper intersection where the two maxima are on top of each other the double-hump intensity distribution shown in the top part of Fig.8. However, in the lower case where the two spots are next to each other we arrive at a single dominant maximum depicted at the lower part of Fig.8.

\begin{figure}
\includegraphics[width=1\linewidth]{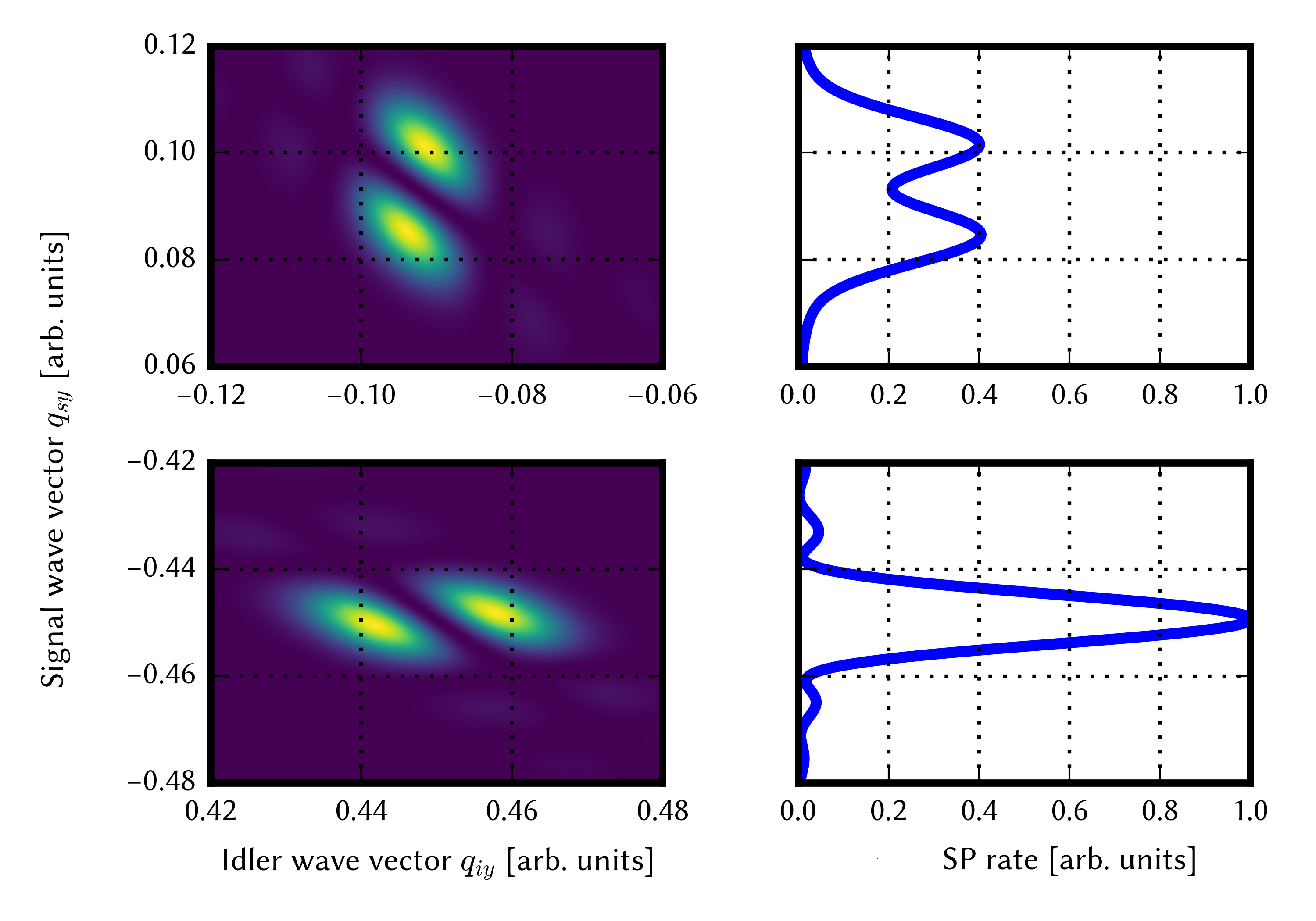}
\caption{Signal photon (SP) count rates (right column) as tomographic cuts through the upper (top) and lower (bottom) intersections (left column) of the phase matching curves depicted in Fig.7. The cuts are along the horizontal axis corresponding to the wave vector component q$_{iy}$ of the idler photon.}\label{fig8}
\end{figure}

\section{Complementarity in action}

In the preceding section we have gained a deeper understanding of the rather intricate mode structure arising from type-II-phase matching in SPDC when pumped by a TEM$_{01}$-mode. This knowledge enables us to perform for the first time a double-slit experiment with single photons in a TEM$_{01}$-like-mode structure while simultaneously obtaining ``which-slit'' information.

Unfortunately, the count rates in our experiment are too small to obtain the statistics necessary to observe the effect. Fortunately, the agreement between simulation and experiment is much better than the experimental error. For this reason, we can base our conclusions on complementarity solely on our theoretical analysis.

\begin{figure}
\includegraphics[width=0.8\linewidth]{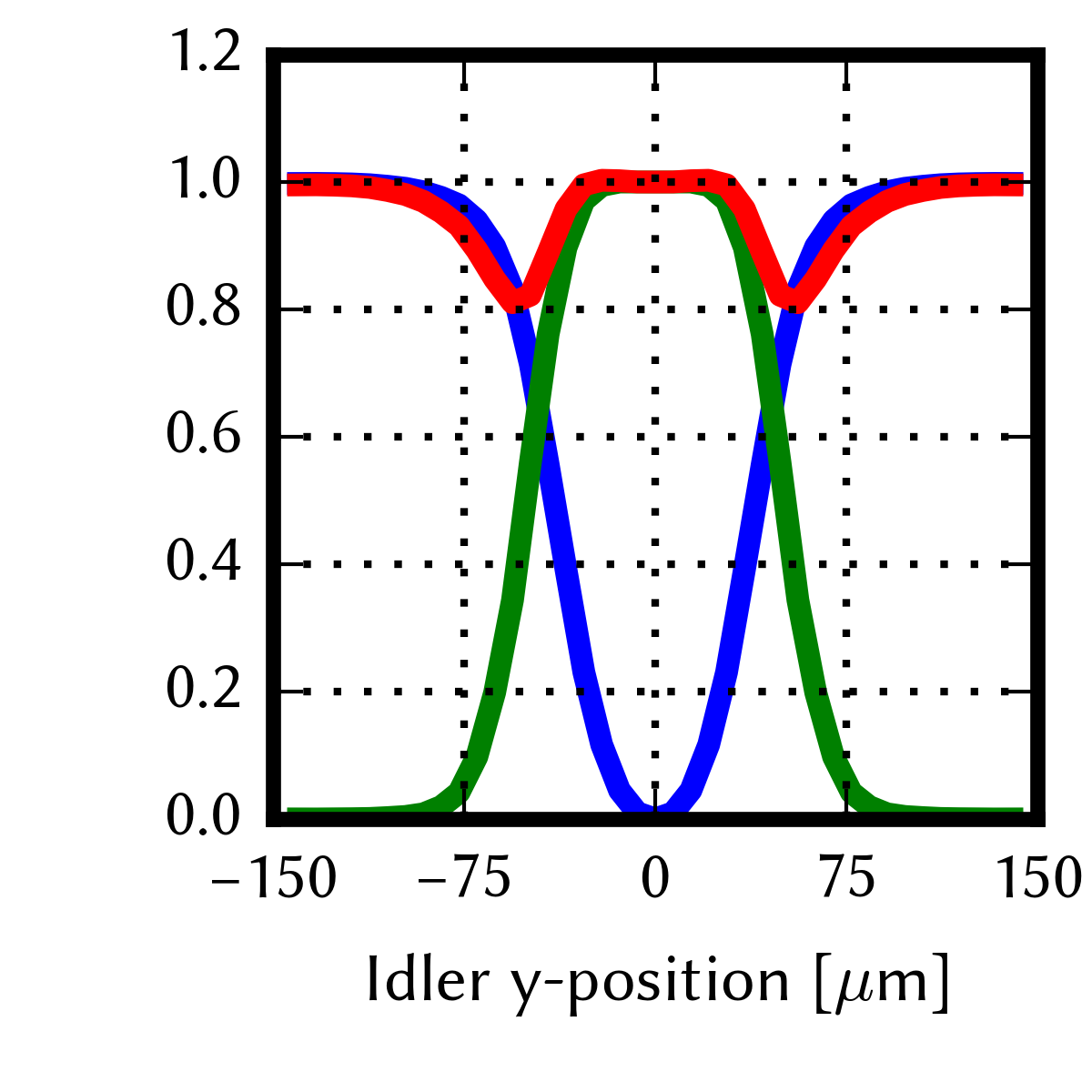}
\caption{Numerical simulation of our two-photon double-slit experiment in the presence of a circular aperture selecting only the double-hump structure at the double-slit. Visibility V (green) measured in the far field, and distinguishability D (blue) obtained in the near field of the signal photons behind the double-slit in coincidence with the idler photons, together with the sum (red) $V^2 + D^2$ as a function of the vertical position of the idler detector. We emphasize that the red curve never reaches above unity, in complete agreement with the inequality, Eq.(1), expressing the principle of complementarity.}\label{fig9}
\end{figure}

In Fig.9 we present a numerical simulation of the visibility V (green), the distinguishability D (blue), and the sum $V^2 + D^2$ (red) in the presence of a circular aperture mapping the TEM$_{01}$-mode onto the double-slit. Three features stand out most clearly: (i) The ``which-slit'' information, that is D, varies from unity to zero, and back to unity when we move the idler detector in near field selecting signal photons across the double-hump structure. (ii) In this process the visibility V goes from zero to unity, and back to zero, and (iii) we find for large domains of positions of the idler detector the equality $V^2 + D^2 = 1$.

The condition for observing high visibility is the non-distinguishability of the photons in the two maxima which occurs when the idler detector is positioned exactly in the middle of the mode. In this case all measured signal photons belong to the TEM$_{01}$-like mode. When the idler detector is displaced from this position not all measured signal photons belong to this mode anymore. It is these photons which provide us with ``which-slit'' information in the near-field correlation measurement.

As a result, the sum of the squares of the visibility and distinguishability is not unity anymore but falls below the bound of the inequality, Eq.(1). This drop of the sum shown in Fig.9 is especially interesting in light of the recent proposal \cite{Quian} of an equality rather than an inequality, which arises by adding the square of the concurrence to the squares of the visibility and distinguishability. Indeed, Ref. \cite{Quian}  argues that the concept of entanglement expressed by the concurrence is the quantity which has been missing so far in our understanding  of the principle of complementarity. Unfortunately, a more detailed analysis of this phenomenon goes beyond the scope of the present article and has therefore to be postponed to a future publication.

\section{Summary}

In summary, we have shown experimentally and by numerical simulations which reproduce all essential features of our experiment that type-II phase matching in SPDC pumped by a TEM$_{01}$-mode produces a rich variety of light structures dominating single-photon interference and coincidence measurements. The two emitted light cones of the signal and the idler photons are asymmetric. Indeed, the \textit{double}-hump structure of the TEM$_{01}$-pump mode appears only on the top of the ring. On the bottom we find a \textit{single}-hump TEM$_{00}$-like intensity distribution.

When we investigate the principle of complementarity with bi-photons created in this way we must avoid this mixture of modes by applying circular apertures which select the appropriate photons. They eliminate those that pollute especially the near-field measurement leading to possibly wrong conclusions.

We conclude by emphasizing that the principle of complementarity in the spatial domain is intimately linked to, and is part and parcel of the detection system. Measuring photons in coherent higher-order modes results in high visibility but in no ``which-slit'' information, and vice versa.

\begin{acknowledgements}
We thank R.W. Boyd, K. Dechoum, B.-G. Englert, L. Happ, M. Hillery, J. Leach, G. Leuchs, P.W. Milonni, M. Efremov, J. Reintjes, M.O. Scully, M.J.A. Sp\"ahn, M. Wilkens, M.S. Zubairy, and S.-Y. Zhu for many fruitful discussions on this topic. WPS is grateful to the Hagler Institute for Advanced Study at  Texas A\&M University for a Faculty Fellowship and to Texas A\&M AgriLife Research. The work of IQ$^{ST}$ is financially supported by the Ministry of Science, Research and Arts Baden-W\"urttemberg. We acknowledge the support of the Deutsche Forschungsgemeinschaft (German Research Foundation) and the Open Access Publication Fund of Potsdam University.\\
\end{acknowledgements}


\begin{thebibliography}{99}
\bibitem{footnote01} We emphasize that in contrast to familiar experiments employing photons from SPDC our arrangement relies on the entanglement in position and momentum variables of the classical mode functions rather than polarization. The latter is only used to separate idler and signal photons with the help of the polarizing beam splitter (PBS) in front of the double-slit shown in Fig.1.
\bibitem{footnote02} We have also employed two additional crystals of lengths 1 mm and 0.5 mm to probe the influence of phase matching.
\end{thebibliography}
\end{document}